\documentstyle[11pt,fleqn,epsfig]{article}

\newcommand{\AmS}{{\protect\the\textfont2
  A\kern-.1667em\lower.5ex\hbox{M}\kern-.125emS}}

\begin{document}
\vspace*{-1.8cm}
\begin{flushright}
\flushright{\bf LAL 99-53}\\
\vspace*{-0.5cm}
\flushright{October 1999}
\end{flushright}
\vskip 2.5 cm

\centerline {\LARGE\bf Determination of $s(x)$ and $\bar{s}(x)$ from a global QCD analysis}
\vspace{2mm}

\vskip 1.5 cm

\begin{center}
{\Large\bf F. Zomer}\\
\end{center}

\begin{center}
{\bf\Large Laboratoire de l'Acc\'el\'erateur Lin\'eaire}\\
{IN2P3-CNRS et Universit\'e de Paris-Sud, BP 34, F-91898 Orsay Cedex}
\end{center}


\begin{abstract}
A new global QCD analysis of DIS data is presented. The 
 $\nu Fe$ and $\bar{\nu}Fe$  differential cross-section data are included
to constrain the strange component of the nucleon sea. As a result we found 
a hard strangeness at high-$x$ and some evidence
 for an asymmetry between $xs(x)$ and $x\bar{s}(x)$. 
\end{abstract}


\section{Introduction}

The starting point of the work presented in this communication is an
 evidence: what is measured  by $\nu$ and $\bar{\nu}$ fixed target
 experiments are the differential cross-sections and not the structure
 functions.
 The former observables are described by a combination of three structure
 functions 
 $F_i^{\nu}$ and $ F_i^{\bar \nu}$ ($i=1,2,3$) and, within pQCD,
  $F_i^{\nu}\ne F_i^{\bar \nu}$. There
 is no magic, to solve a system 2 equations with 6 unknowns one must provide
 some extra informations ...
 And the main part of these informations comes from the 
 $W^++s\rightarrow c$ (plus higher order corrections) process: the deuce
 of structure functions [$F_2$, $F_3$], 
 extracted from the mixed $\nu$ and $\bar{\nu}$ data,
 received some large corrections and especially from the strange density
\cite{seligman}. 

 Therefore, unlike what is currently done in the global QCD fits
 \cite{global-fit}, we do not consider here the {\it deuce} [$F_2$, $F_3$]
 but rather all the available differential cross-section measurements:
 BEBC(H and D targets), CDHS (H target) and CDHSW (Fe target)\cite{data}.
 The fit presented here (see \cite{nous} for the details)
 is very similar to the CTEQ5F3 one \cite{global-fit}.
 The same combinations of parton densities are parametrised at 
 $Q_0=2$GeV and evolved using the DGLAP equations \cite{dglap}.
 In addition to the neutrino data sets, the
 data entering the fit are
 \cite{data}: the fixed target and H1 charged lepton ($\ell^pm$) 
 beam $F_2$; the Drell-Yan
 differential cross-sections and asymmetries. To avoid higher-twist
 corrections we also apply some data rejection cuts: $Q^2\ge 3.5$GeV$^2$ 
 and $W^2\ge 10$GeV$^2$.

 From this fit we can
 {\it i)} determine $xs(x)$ and $x\bar s(x)$ within an inclusive analysis, 
 {\it ii)} test the compatibility between charged lepton and neutrino beam 
 observables (see \cite{seligman} where an incompatibility is reported).
 In this communication we concentrate on the strangeness and we shall thus 
start by giving, in section 1,
 some details concerning the nuclear corrections applied to the CDHSW data.
This is an important feature since
 this data sample is the most significant statistically.
The fit results concerning the two items {\it i)} and {\it ii)} pointed
 out above are given in section 2.

\section{On the nuclear corrections applied to $Fe$ target data}
\begin{figure}[htb]
\centering
\includegraphics*[width=90mm]{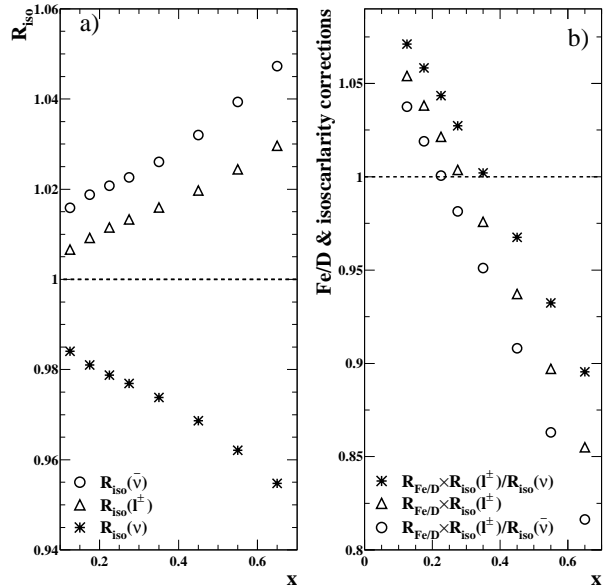}
\caption{Nuclear correction applied to CDHSW data (see text).}
\label{isocor}
\end{figure}

All CDHSW  data are 
obtained from scattering off Iron nuclei. Since the 
theoretical understanding of nuclear effects in 
heavy nuclei is still uncertain and model dependent
\cite{michele}, we adopt an empirical 
procedure to perform the nuclear corrections. The basis of the procedure
is that, in the `naive' QPM,
 $\ell^\pm$ and neutrino nucleon structure functions are proportional at 
high-$x>\approx 0.1$.
The experimental $\nu(\bar{\nu}) Fe$ differential cross-sections 
are then fitted to 
$
d\sigma^{\nu(\bar{\nu})Fe}= 
{d\sigma_{iso}^{\nu(\bar{\nu})Fe}}/{R_{iso}^{\nu(\bar{\nu})}}
$, 
where $R_{iso}^{\nu(\bar{\nu})}$ is the correction factor for the 
non-isoscalarity of Iron
\begin{equation}\label{iso}
R_{iso}^{\nu(\bar{\nu})} =  
\frac{(d\sigma^{\nu(\bar{\nu})p}
+ d\sigma^{\nu(\bar{\nu})n})/2}{(Z \, d\sigma^{\nu(\bar{\nu})p} + 
(A-Z) \, d\sigma^{\nu(\bar{\nu})n})/A}
\end{equation}
and $d\sigma_{iso}^{\nu(\bar{\nu})Fe}= 
d\sigma^{\nu(\bar{\nu})D} 
\cdot 
R_{nucl}^{iso}$ with
 $R_{nucl}^{iso} = R_{Fe/D} \cdot R_{iso}^{\ell}$.  
  The first factor of the last equation is the $Fe/D$ structure
 function ratio 
$R_{Fe/D} = {F_2^{\ell^\pm Fe}}/{F_2^{\ell^\pm D}} $
which is obtained 
from a fit to the most precise    
experimental data on 
$F_2^{\ell^\pm Fe}/F_2^{\ell^\pm D}$ \cite{emc}, uncorrected for isoscalarity 
(the $x$ range is $[0.1,0.65]$).  
The second factor contains the isoscalarity corrections and is
 computed from eq. (\ref{iso}) using $F^{\ell^\pm p,n}_2$. The different 
 contributions to the nuclear corrections are shown in fig. \ref{isocor}. 
 This figure shows that 
 isoscalar correction induces some large differences between the $\nu$
 and $\bar \nu$ data at high-$x$. It is to be noticed that these corrections
 are also applied to the CCFR [$F_2$, $F_3$] data \cite{seligman} used 
 actually to determined the valence quarks in the global fits \cite{global-fit}.
\section{Results}
Two fits were performed. In {\tt fit1} we fixed $s=\bar s$ but, unlike
 what is done in ref. \cite{global-fit}, we parametrise $s$ independently of 
the non strange sea. In {\tt fit2}
 $s$ and $\bar s$ are parametrised independently requiring 
 $\int_0^1(s(x)-\bar s(x))dx=0$. Of course, other parton densities are 
parametrised in order to determined all the observables entering the fits.

As a first result we indicate that from a statistical point of view a perfect 
agreement between $\ell^\pm$ and $\nu(\bar{\nu})$ beam observables is
 observed with both fits. The disagreement observed in \cite{seligman} is
 therefore most likely related to extraction procedure of the
 deuce [$F_2$, $F_3$].
 A good agreement is also found between the BEBC $\nu(\bar{\nu})$ D target
data and the CDHSW data. This gives us some confidence in the large 
nuclear corrections described in the previous section. 

\begin{figure}[htb]
\vspace{-1pt}
\centering
\includegraphics*[width=80mm]{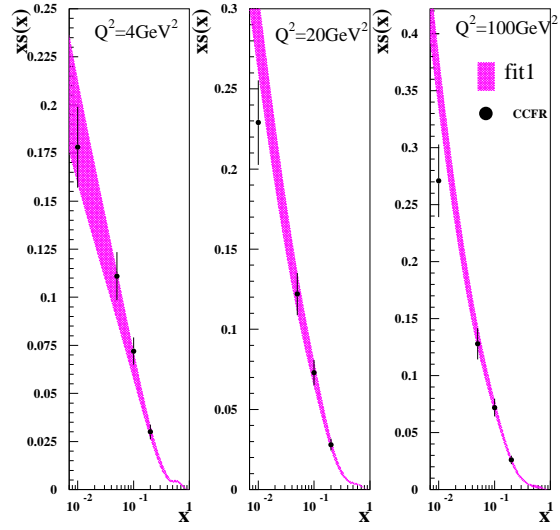}
\caption{Results of {\tt fit1} (see text).}
\label{splots}
\end{figure}

 The strange density determined by {\tt fit1} is shown in fig. \ref{splots}
 where the error bands include the propagation of all the experimental
 uncertainties. A good agreement is found with the CCFR di-muon analysis
 \cite{dimuon} results and one can also remark that $xs(x)$ is `hard' at
 $x>0.4$. Notices that a worse $\chi^2$ (significantly worse) is obtained if,
 as in ref. \cite{global-fit}, one fixes
 $ s=\bar s=(\bar u+\bar d)/4$ in {\tt fit1}.
\vspace{-0.5cm}
\def \textfraction{0}

\begin{figure}[hb]
\vspace{0pt}
\centering
\includegraphics*[width=80mm]{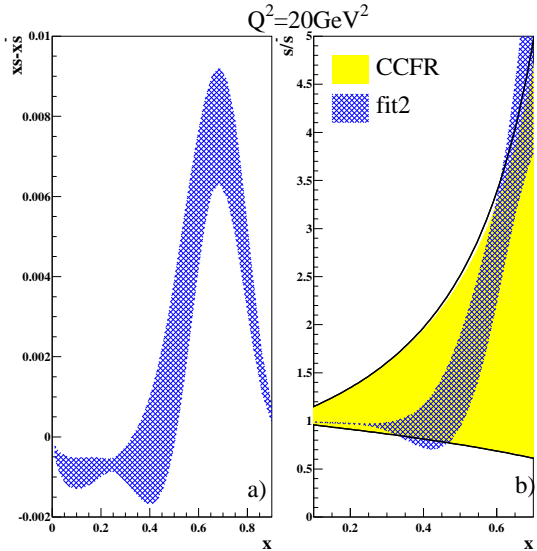}
\caption{Results of {\tt fit2} (see text).}
\label{s_over_sbar}
\end{figure}

The results of {\tt fit2} for $xs(x)-x\bar s(x)$ and $s(x)/\bar s(x)$ are
 shown in fig. \ref{s_over_sbar}.
\def \textfraction {0.2}
\newpage

\begin{figure}[htb]
\vspace{0pt}
\centering
\includegraphics*[width=85mm]{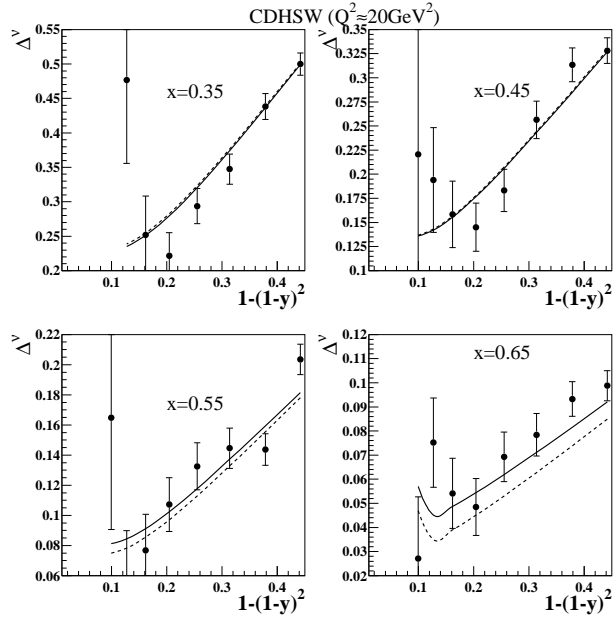}
\vspace{0.8cm}
\caption{ Comparison between  {\tt fit1} (hashed line) and {\tt fit2}
 (full line) (see text).}
\label{cdhsw_diff_18gev}
\end{figure}

 A significant non-singlet
 $x(s(x)-\bar s(x))$
 component is observed. From {\tt fit1} to {\tt fit2} the contribution of the
  $\nu Fe$ data to the global $\chi^2$ decreases. Our results for 
$s(x)/\bar s(x)$ are again not incompatible with the determination
 of CCFR \cite{dimuon}.
 In fig. \ref{cdhsw_diff_18gev} the following observable
 $$
\Delta^{\nu- \bar \nu}\equiv \frac{4\pi x(M_W^2+Q^2)^2}{G_F^2  M_W^4}
 \biggl[
\frac{d^2\sigma^{\nu N}}{dxdQ^2}-\frac{d^2\sigma^{\bar{\nu} N}}{dxdQ^2}
\biggr]\,.
$$
 is compared to 
{\tt fit1} and {\tt fit2}.  In the QPM one has 
 $\Delta^{\nu- \bar \nu} 
\propto x s(x) - x \bar{s}(x)+Y_-[xu_v(x)+xd_v(x)]$ so that
 fig. \ref{cdhsw_diff_18gev} clearly demonstrates the sensitivity of
  the inclusive $\nu(\bar{\nu})$ data to the strange quarks.
\vspace{0.5cm}
\vspace{0.8cm}
\section{Summary}
\vspace{0.2cm}
We have shown that if the $\nu(\bar{\nu})$ differential cross-section 
enters the pQCD fits instead of the neutrino structure functions one obtains:
 {\it i)} a good agreement between $\ell^\pm$ and $\nu(\bar{\nu})$ DIS 
 observables; {\it ii)} a constraint on $xs$ and $x\bar s$. However, since the
 nuclear correction applied to the $\nu(\bar{\nu}) Fe$ data are large and
 determine empiricaly our results may be taken at a qualitative level: 
 we have obtained some `hard' $xs$ and $x\bar s$ - with  $xs$ `harder'
 than $x\bar s$ - 
distributions at high-$x$ and a non vanishing non-singlet density
 $x(s-\bar s)$.

 Finally, let us emphasize 
 that a more quantitative determination of the strangeness
  may be possible in a near future 
 using the forthcoming $\nu(\bar{\nu}) Fe$ CCFR
 (see U.K. Yang's contribution) and 
  $\nu(\bar{\nu}) Pb$ CHORUS (see R. Oldeman's contribution)
cross-section measurements.

\end{document}